\documentclass[12pt]{article}
\usepackage{epsfig}

\usepackage{float}
\usepackage{graphicx}

\begin{document}
\rightline{NKU-2016-SF4}
\bigskip

\newcommand{\be}{\begin{equation}}
\newcommand{\ee}{\end{equation}}
\newcommand{\noi}{\noindent}
\newcommand{\ra}{\rightarrow}
\newcommand{\bib}{\bibitem}
\newcommand{\refb}[1]{(\ref{#1})}

\newcommand{\bff}{\begin{figure}}
\newcommand{\eff}{\end{figure}}

\begin{center}
{\Large\bf P-V criticality in AdS  black holes of  massive gravity}

\end{center}
\hspace{0.4cm}
\begin{center}
Sharmanthie Fernando \footnote{fernando@nku.edu}\\
{\small\it Department of Physics, Geology \& Engineering Technology}\\
{\small\it Northern Kentucky University}\\
{\small\it Highland Heights}\\
{\small\it Kentucky 41099}\\
{\small\it U.S.A.}\\

\end{center}

\begin{center}
{\bf Abstract}
\end{center}

In this paper we have studied the extended phase space thermodynamics in the canonical ensemble of  black holes in massive gravity in AdS space. The black holes considered here belong to  a theory of massive gravity where the graviton gain a mass due to Lorentz symmetry breaking. We have computed various thermodynamical quantities such as temperature, pressure, Gibbs free energy and specific heat capacity. The local and the global thermodynamical stability of the black holes are studied in detail. For a specific value of the  parameter in the theory, the black holes undergo a first order phase transition similar to the Van der Waals phase transitions between gas and liquid under constant temperature. This transition is between the small and the large black holes. The critical exponents are computed at the critical values and shown to be the same as for the Van der Waals critical exponents.

\hspace{0.7cm}


\section{ Introduction}

Black hole thermodynamics in anti-de Sitter space  have attracted lot of attention in the recent past due to many reason: one of them is the seminal work of Hawking and Page \cite{page}; there, Hawking and Page discovered a first  order phase transition between the Schwarzschild-anti-de Sitter black hole and the thermal anti-de Sitter space. This phenomenon, well known as  the Hawking-Page transition is explained as the gravitational dual of the QCD confinement/de-confinement transition by Witten \cite{witten1} \cite{witten2}. Another landmark in thermodynamics of AdS black holes was the discovery of phase transitions similar to Van der Waals liquid/gas transitions in Reissner-Nordstrom AdS black holes by Chamblin et.al \cite{cham1} \cite{cham2}. They studied the Reissner-Nordstrom AdS black holes in both canonical and well as grand canonical ensemble and discovered a first order phase transition between small and large black holes.  Fernando \cite{fernando1} extended this idea by studying the  Born-Infeld-AdS black hole  in the grand canonical ensemble to discover a first order phase transition there too.

Another reason studies of AdS black holes have take center stage is due to the AdS/CFT correspondence \cite{juan}.

Current interest on studies of thermodynamics of AdS black holes have increased due to the rich structures found by treating the cosmological constant as a thermodynamical variable. The first law of black holes is modified by including a $V dP$ term where the pressure $P$ is given by $-\Lambda/ 8 \pi$. In this scenario, the mass of the black hole $M$ is considered as the enthalpy of the black hole system instead of the internal energy \cite{kastor} \cite{dolan}. One of the first works to explore the extended phase space thermodynamics in AdS black holes was the paper by Kubiznak and Mann \cite{mann}. There the critical behavior of the charged black hole in 4 dimensions was studied in detail; the critical exponents were shown to be same as for the Van der Waals liquid/ gas system. Since then there have been large number of papers devoted in studying interesting properties of black holes in the extended phase space. Reetrant phase transitions and van der Waals behavior for hairy black holes were studied in \cite{mann2}. Thermodynamics of rotating black holes and black rings were studied in \cite{mann3}. Charged rotating black holes and Born-Infeld-AdS black holes were studied in \cite{mann4}. In an interesting paper, the geometry of a black hole which behaves exactly like the Van der Waals black holes were presented by Rajagopal et.al \cite{mann5}.  There are many other works related to this concept including \cite{cao}  \cite{liu2} \cite{hendi} \cite{hendi2} \cite{mo} \cite{zhang} \cite{li} \cite{mo2} \cite{cai} \cite{cai2} \cite{azg}\cite{pou}.

Massive gravity theories where the graviton acquire a mass has become increasingly popular in the current literature. One of the reasons for this popularity is that this modification of general relativity in principle could explain the acceleration of the universe without introducing the component of ``dark energy.'' Current experimental observations put constraints on the mass of the graviton; recent observation of gravitational waves by Advanced LIGO has put constants on the mass of the  $m_g < 1.2 \times 10^{-22} eV/c^2$ \cite{ligo}.

Fierz and Pauli in 1939 \cite{pauli} are the first to develop a theory to include a mass to the graviton. They developed a Lorentz invariant massive spin 2 theory described by a quadratic action. Fierz-Pauli theory has five degrees of freedom. One of the problems of this theory was the appearance of the so called  Boulware and Deser ghosts \cite{boul}. The problem of ghosts were recently solved by de Rham, Gabadadz and Tolley  with a new massive gravity theory  \cite{drgt1}\cite{drgt2}. Two other models for massive gravity which are free from ghosts are the DGP model \cite{dgp} and the ``new massive gravity theory''   in three dimensions \cite{town}. There are large volume of   works related to massive gravity in the literature and there is no space to present all here; instead we will direct the reader to  two excellent reviews on the topic by de Rham \cite{claudia} and  Hinterbichler \cite{kurt}.

Out of the many massive gravity theories, the one we will consider in this paper is  a theory with Lorentz symmetry breaking by a space-time dependent condensates of scalar fields. Such scalar fields act as  Goldstone fields and are coupled to gravity via non-derivative coupling.

When Lorentz symmetry is broken spontaneously, the graviton acquire a mass very similar to the Higgs mechanism.  A  review of Lorentz violating massive gravity theory can be found in \cite{dubo} \cite{ruba2}.

The action for this theory is given by,
\be \label{action}
S = \int d^4 x \sqrt{-g } \left[ - M_{pl}^2 \mathcal{R}  + \Omega^4 \mathcal{F}( X, W^{ij}) \right]
\ee
Here the first term is the  Einstein-Hilbert Lagrangian for general relativity and $\mathcal{R}$ is the scalar curvature of the space-time geometry. The second term $\mathcal{F}$ composed of two functions $X$ and $W$ defined in terms of the scalar fields as,
\be \label{scalar1}
X = \frac{\partial^{\mu} \Phi^0 \partial_{\mu} \Phi^0 }{ \Omega^4}
\ee
\be \label{scalar2}
W^{i j} = \frac{\partial^{\mu} \Phi^i \partial_{\mu} \Phi^j}{ \Omega^4} - \frac{\partial^{\mu} \Phi^i \partial_{\mu} \Phi^0  \partial^{\nu} \Phi^j \partial_{\nu} \Phi^0 }{ \Omega^8 X}
\ee
$\Omega$ has dimensions of mass and is in the order of $ \sqrt{ m_g M_{pl}}$: here  $m_g$  is the graviton mass and $M_{pl}$ the Plank mass \cite{dubo} \cite{luty} \cite{ruba} \cite{pilo1}. The scalar fields $\Phi^0, \Phi^i$  are responsible for breaking Lorentz symmetry spontaneously when  they  acquire a vacuum expectation value.

The paper is organized as follows: in section 2, the black hole in massive gravity is introduced. Thermodynamics, first law and  Smarr formula  were discussed in section 3. Specific heat capacity is analyzed in section 4. Gibbs free energy is discussed in section 5. Finally the conclusion is given in section 6.


\section{ Introduction to AdS black holes in massive gravity}

In this section we will present  AdS black holes in massive gravity. The metric of such black holes were derived in detail in   \cite{tinya} and \cite{pilo}.  When the equations of motion for the action in eq$\refb{action}$ is written, it leads to complex set of non-linear equations. It is impossible to solve those equations for a generic function  $\mathcal{F}$. Hence, in \cite{tinya}, the function $\mathcal{F}$ was chosen in such a way that the resulting equations can be solved analytically. Therefore the function $\mathcal{F}$ of the solution given in this paper is given by,

\be
\mathcal{F} = \frac{ 12 b^6} { \lambda} \left( \frac{ 1}{ X} + \chi_1 \right) - \left( \chi_1^3 - 3 \chi_1 \chi_2 - 6 \chi_1 + 2 \chi_3 - 12\right)
\ee
where,
\be
\chi_n = Tr( W^n)
\ee

The scalar field $\Phi$  in eq$\refb{scalar1}$ and eq$\refb{scalar2}$ for this particular solution is given by,
\be \label{scalar}
\Phi^0 = \Omega^2 \left( t + \beta(r) \right); \hspace{1 cm} \Phi^i =   \Omega^2 b x^i
\ee
where
\be \label{beta}
\beta(r) = \pm \int \frac{ dr} { h(r)} \left[ 1 - h(r) \left( \frac{ \gamma Q^2 \lambda(\lambda-1)}{ 12 m_g^2 b^6}\frac{1}{ r^{\lambda+2}} + 1 \right)^{-1} \right]^{1/2}
\ee
The parameter $Q$ represents a scalar charge related to massive gravity and $\gamma = \pm 1$. The constant $\lambda$ in eq$\refb{beta}$ is an integration constant and is positive. In the eq$\refb{beta}$, $m_g$ is the mass of the graviton. The parameter $b$ in eq$\refb{beta}$ and eq$\refb{scalar}$ is an intergration constant in the theory.

For the above function $\mathcal{F}$, the solution for the metric is given by,
\begin{equation} \label{metric}
ds^2 = - h(r) dt^2 + \frac{ dr^2}{ h(r)} + r^2 ( d \theta^2 + sin^2 \theta d \phi^2)
\end{equation}
where,
\begin{equation} \label{hr}
h(r) = 1 - \frac{ 2 M} { r} - \gamma \frac{ Q^2}{r^{\lambda}} - \frac {\Lambda r^2}{3}
\end{equation}
Here, the cosmological constant $\Lambda $ is related to the constant $b$ and $m_g$ as, $\Lambda = 2 m_g^2 ( 1 - b^6)$. Note that when the above black hole solutions were derived by Bebronne and Tinyakov\cite{tinya}, the cosmological constant term $-\frac{ \Lambda r^2}{3}$ was not in the function $h(r)$ since $b$ was chosen to be $1$. However, it is still possible to choose $ b \neq 1$ so that there will be a non-zero cosmological constant term in the metric.  
Hence here, we will choose $ b >1$ so that  $\Lambda <0$ term to be included in the metric.

When $\lambda  <1$, the term $\frac{ - \gamma Q^2}{r^{\lambda}}$ in $h(r)$ dominates at large distances and the ADM mass of such solutions become divergent. When $\lambda >1$, for large distances the metric approaches the usual Schwarzschild-AdS black hole with a finite mass $M$; we will choose $\lambda >1$ in the rest of the paper. \\

\noi
When $ \gamma = 1$, the geometry of the black hole is  very similar to the Schwarzschild-AdS  black hole with a single horizon. When $\gamma = -1$, the geometry is similar to the well known  Reissner-Nordstrom-AdS  charged black hole. In this case here could be two horizons.

There are several works related to the black holes in Lorentz breaking massive gravity discussed above. Phase transitions of non-extended phase space of AdS massive gravity black holes were discussed by Fernando in \cite{fernando1}. Scalar and Dirac field perturbations of the black hole with $\Lambda =0$ were studied in  \cite{fernando2}\cite{fernando3}. Thermodynamics and phase transitions for $\Lambda=0$ case were studied in  \cite{capela} \cite{mirza}.  Stability of spherically symmetric solutions of Lorentz breaking massive gravity was studied in \cite{and}.


\section{Thermodynamics, first law and the Smarr formula}

In this section we will  derive the thermodynamical quantities, first law  and presented the Smarr formula for the black hole in massive gravity.


\subsection{ Defining the thermodynamical quantities}

\begin{figure} [H]
\begin{center}
\includegraphics{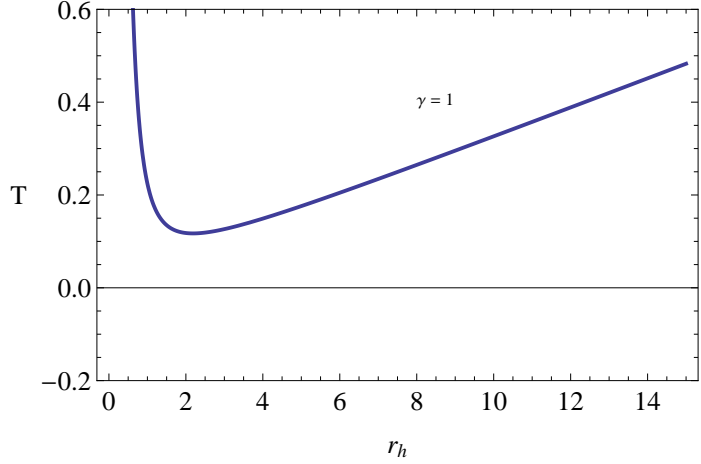}
\caption{The figure shows  $T$ vs $r_h$ for $ \gamma = 1$ for varying temperature. Here  $\lambda = 1.955, P = 0.01592$, and $ Q = 1.22$.}
\label{temp1}
 \end{center}
 \end{figure}

\begin{figure} [H]
\begin{center}
\includegraphics{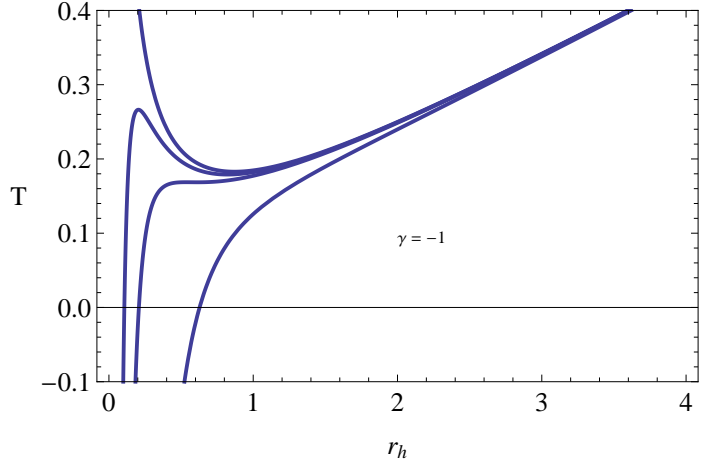}
\caption{The figure shows  $T$ vs $r_h$ for $ \gamma =-1$ for varying temperature. Here  $\lambda = 1.55$, and $ P =  0.0525$. The top graph is at $Q=0$ and the rest have $Q = 0.24, 0.41, 1.16$}
\label{temp2}
 \end{center}
 \end{figure}

The Hawking temperature of the black hole is given by,
 \be \label{temp}
 T_H =  \frac{ 1}{ 4 \pi}  \left| \frac{ dh(r)}{ dr} \right|_ { r = r_h} = \frac{1}{ 4 \pi} \left(\frac{ 2 M}{ r_h^2} + \frac{ \gamma Q^2 \lambda}{ r_h^{ \lambda + 1}}   - \frac{ 2 \Lambda r_h}{ 3}\right)
 \ee
 Here $r_h$ is the black hole event horizon. Since at the black horizon $ h(r_h) = 0$, the mass of the black hole could be written as,
 \be \label{mass}
 M = \frac{r_h}{2} - \frac{ \gamma Q^2}{ 2 r_h^{( \lambda -1)} }-  \frac{ r_h^3 \Lambda}{6}
 \ee
 The value of the mass $M$ can be substituted to the temperature in eq$\refb{temp}$ and rewrite it as,
 \be
T = \frac{ 1 }{ 4 \pi} \left( \frac{1}{ r_h}  -  r_h \Lambda   + \frac{ \gamma  ( \lambda -1)Q^2}{ r_h^{ \lambda +1} } \right)
\ee
The temperature is plotted in Fig$\refb{temp2}$ and Fig$\refb{temp1}$. When $\gamma =1$, the temperature has a minimum. Below this minimum black holes cannot exist for a given $\Lambda$ value. For $\gamma =-1$, the temperature could have an infection point for a given value of $Q$ and $\Lambda$.

The entropy of the black hole is given by the area law, $ S = \pi r_h^2$. The entropy is the conjugate quantity to the temperature.  In the extended phase space, the cosmological constant is treated as the thermodynamic pressure with the relation,
\be
P = -\frac{ \Lambda}{ 8 \pi}
\ee
The corresponding conjugate quantity, the volume $V$ is given by, $\frac{ 4 \pi r_h^3}{3}$. The scalar potential corresponding to the scalar charge $Q$ is given by,
\be
\Phi  = - \frac { \gamma Q}{ r_h^{ \lambda -1}}
\ee

\subsection{ First law and the Smarr formula  in the extended phase space}
 
 The first law of the black hole  in consideration with the thermodynamic quantities defined in the previous section is given by
 \be \label{flaw}
 dM = T dS + \Phi dQ + V dP
 \ee
 Notice that in this case $M$ is not the internal energy $E$. Instead, one has to treat $M$ as the enthalpy $H$ given by $ E + P V$ so that the first law is valid. From the first law, all the thermodynamic variables defined above could be obtained as follows:
\be 
\left ( \frac{ \partial M} { \partial S } \right) _{Q,P} =  T
\ee
\be 
\left ( \frac{ \partial M} { \partial Q } \right) _{S,P} = \Phi
\ee
 \be 
\left ( \frac{ \partial M} { \partial P } \right) _{S,Q} =  V
\ee
 One can combine the thermodynamical quantities presented above to obtain the Smarr formula as,
 \be
 M = 2 T S + \frac{\lambda}{2} \Phi Q - 2 P V
 \ee
The Smarr formula also can be obtained using the scaling argument presented by Kastor et.al \cite{kastor}. Notice that when $\lambda =2$, the Smarr formula simplifies to the one for the Reissner-Nordstrom-AdS black hole obtained by  Kubiznak and Mann in \cite{mann}.


\subsection{ Behavior of pressure}

\begin{figure} [H]
\begin{center}
\includegraphics{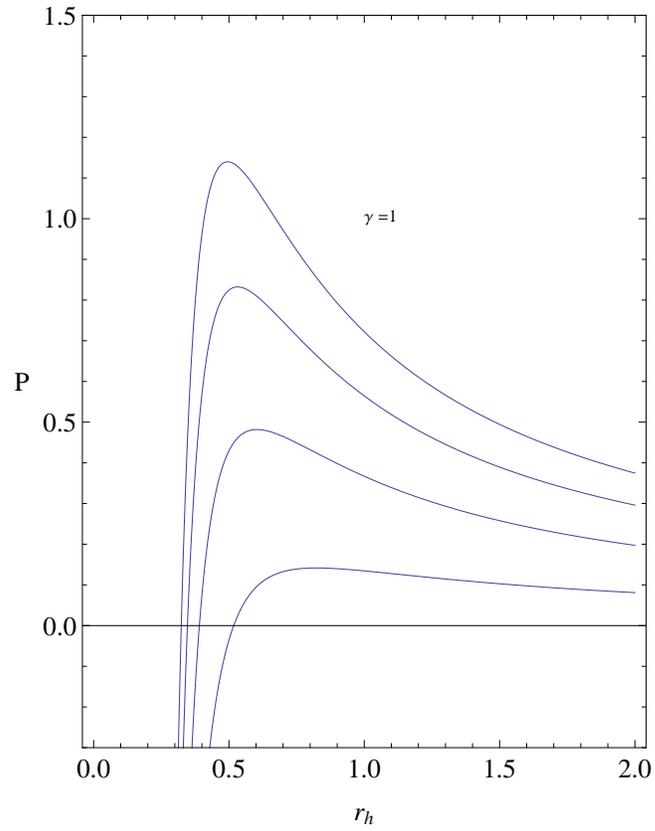}
\caption{The figure shows  $P$ vs $r_h$ for $ \gamma =1$ for varying temperature. Here  $\lambda = 2.865$, and $ Q = 0.334$. For large temperature the peak is higher.}
\label{pvsr2}
 \end{center}
 \end{figure}

For a fixed scalar charge $Q$, one can substitute $\Lambda = - 8 \pi P$ in eq$\refb{temp}$ to obtain $P = P(V,T)$ as,
\be
P = \frac{ T}{ 2 r_h} - \frac{ 1}{ 8 \pi r_h^2}  + \frac{ Q^2 \gamma ( 1 - \lambda)}{ 8 \pi r_h^{ 2 + \lambda}}
\ee
Also, the black hole radius $r_h$ is given by,
\be
r_h = \left(\frac{ 3 V} {4 \pi}\right)^{1/3}
\ee
When $P$ is plotted vs $r_h$ for both values of $\gamma$, the behavior is quite different.  For $\gamma=1$, there are no critical points as demonstrated in Fig$\refb{pvsr2}$. There is  maximum value of pressure $P_{max}$ for  given temperature $T$. At that point $\frac{\partial P}{ \partial r} =0$. One can solve the equation to obtain,
\be \label{pmax}
P_{max} = \frac{ 4 \pi r_h T ( 1 + \lambda) - \lambda}{ 8 \pi r_h^2 ( 2 + \lambda)}
\ee

\begin{figure} [H]
\begin{center}
\includegraphics{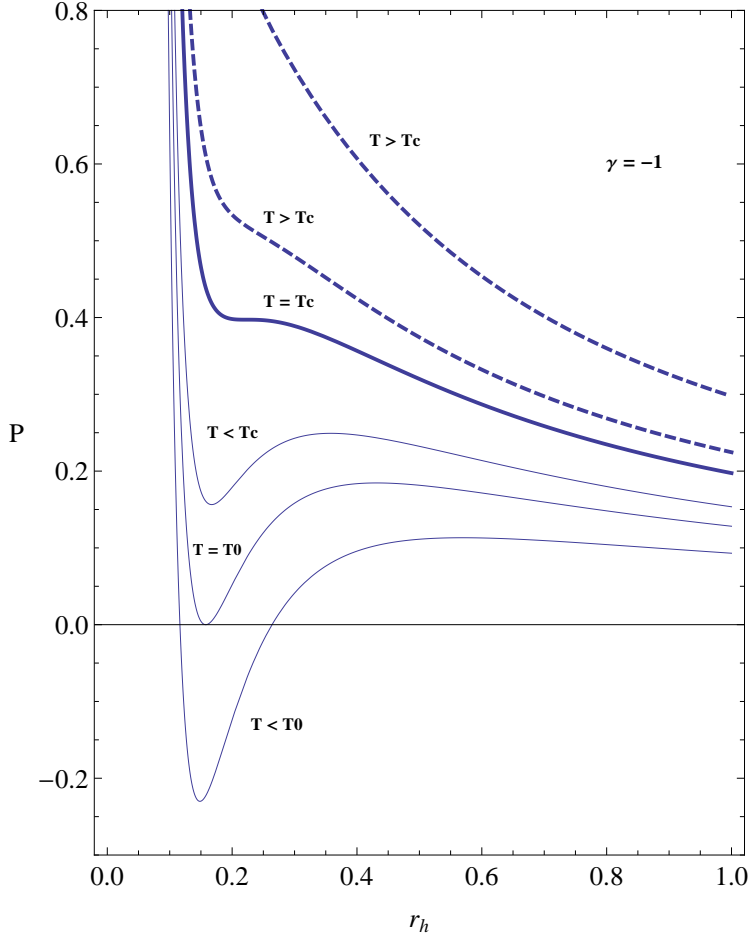}
\caption{The figure shows  $P$ vs $r_h$ for $ \gamma = -1$ for varying temperature. Here  $\lambda = 1.98$, and $ Q = 0.094$.}
\label{pvsr1}
 \end{center}
 \end{figure}

For $\gamma=-1$, the behavior is quite different from what is of $\gamma =1$. There are critical points as demonstrated from Fig$\refb{pvsr1}$.

\subsubsection{Critical values and the law of corresponding states for $\gamma =-1$}

By identifying the specific volume $v = 2 r_h$, the equation of state can be rewritten as,
\be
P = -\frac{1}{ 2 \pi v^2} + \frac{T}{ v}  + \frac{Q^2 ( \lambda -1) 2^{\lambda}}{ 2 \pi v^{ 2 + \lambda}}
\ee
Since $v \propto r_h$, the graph $P$ vs $v$ will be similar in shape and characteristics to the one of $P$ vs $r_h$. 
The critical temperature $T_c$ is the value of $T$ at which the $P$ vs $v$ curve has an inflection point. At this point,
\be
\frac{ \partial P}{ \partial v } =  \frac{ \partial^2 P}{ \partial v^2 } =0
\ee
After some algebra, $P_c, v_c$ and corresponding $T_c$ can be derived to be,
\be
v_c =  2 r_c =  \left( Q^2 ( \lambda^2 - 1) ( \lambda +2 ) 2 ^{\lambda -1}\right)^{1/\lambda}
\ee
\be \label{tc}
T_c = \frac{ \lambda}{ \pi ( \lambda +1)}  \left( Q^2 ( \lambda^2 - 1) ( \lambda +2 ) 2 ^{\lambda -1}\right)^{-1/\lambda}
\ee
\be \label{pc}
P_c = \frac{ \lambda}{ 2 \pi ( \lambda +2)}  \left( Q^2 ( \lambda^2 - 1) ( \lambda +2 ) 2 ^{\lambda -1}\right)^{-2/\lambda}
\ee
The critical thermodynamic volume is given by,
\be \label{vc}
V_c = \frac{ 4 \pi r_c^3}{ 3}
\ee
Hence $r_c$ and $S_c$ are given by,
\be \label{rc}
r_c = \frac{v_c}{2} =\frac{1}{2} \left( Q^2 ( \lambda^2 - 1) ( \lambda +2 ) 2 ^{\lambda -1}\right)^{1/\lambda}
\ee
\be \label{sc}
S_c = \pi r_c^2  =\frac{\pi}{4} \left( Q^2 ( \lambda^2 - 1) ( \lambda +2 ) 2 ^{\lambda -1}\right)^{2/\lambda}
\ee
The ratio between $P_c, V_c$ ad $T_c$ is given by,
\be
\frac{P_c V_c}{ T_c} = \frac{ (\lambda + 1)}{ 2 ( \lambda + 2)}
\ee
When $\lambda =2$, the above ratio simplifies to $\frac{P_c V_c} { T_c} = 3/8$ which is the same value for the Van der Waals gas-liquid system and also for the RNAdS black hole \cite{mann}.

When $T < T_c$, there can be three branches of black hole solutions (small, intermediate and large). In this case, there is a first oder phase transition of the van der Waals type between the small and the large black holes. Similar to the Van der Waals system \cite{hill}, there is a temperature $T_0$ where the pressure is zero. At this point,
\be
\frac{ \partial P}{ \partial v} = 0; P =0
\ee
The above equations can be solved to yield,
\be
T_0 = \frac{ \lambda}{ 4 \pi ( \lambda +1)} \left( Q^2 ( \lambda^2 -1) \right)^{-1/\lambda}
\ee
\be
v_0 = 2 \left( ( \lambda^2 - 1) Q^2 \right)^{1/\lambda}
\ee

 We can define new parameters as,
 \be 
 p = \frac{P}{P_c}; \hspace{0.5 cm} \nu = \frac{ v}{ v_c}; \hspace{0.5 cm} \tau = \frac{T}{T_c}
 \ee
 where $T_c$, $P_c$ and $v_c$ are given by eq.$\refb{tc}$, eq$\refb{pc}$ and  eq$\refb{vc}$.
 With the newly defined quantities, the equation of state simplifies to a `law of corresponding states', 
 \be
 \tau = p \nu \left(\frac{ \lambda + 1}{ 2 ( 2 + \lambda)}\right) - \left(\frac{ 1}{ \lambda( 2 + \lambda)}\right) \frac{1}{\nu^{( 1 + \lambda)} }+ \left(\frac{ 1 + \lambda}{ 2 \lambda}\right)\frac{1}{ \nu}
 \ee

Notice that when $\lambda =2$, the law of corresponding states given above is exactly the same as for the RNAdS black hole \cite{mann}. 
 
 
\subsubsection{ Critical exponents  for $\gamma =-1$}

In this section we will investigate the critical exponents which describes the behavior of  the physical quantities near the critical points discussed in the previous section. For an excellent introduction on the critical exponents, please refer to the book by Goldenfeld \cite{book1}. First, we will introduce new parameters as follows:
\be
t = \frac{ T}{T_c} -1 = \tau -1; \hspace{0.4 cm} \epsilon = \frac{V}{V_c} -1; \hspace{0.4 cm} \sigma = \frac{v}{v_c}; \hspace{0.4 cm} \epsilon = \sigma^3 -1
\ee
There are four critical components, $\alpha, \beta, \xi, \delta$ which are defined as follows \cite{book1}:\\

\noi

\noi
Critical exponents $\alpha$: $\alpha$ governs the specific heat at constant volume as,
\be
C_V = T \left.  \frac{\partial S}{ \partial T}\right|_V \propto |t|^{-\alpha}
\ee
\noi
Critical exponent $\beta$: $\beta$ governs the quantity $\eta = v_g - v_l$ with the relation,
\be
\eta = v_g - v_l \propto |t|^{\beta}
\ee
\noi
Critical exponent $\xi$: $\xi$ governs the behavior of isothermal compressibility $\kappa_T$ defined as,
\be
\kappa_T = - \left.\frac{1}{V} \frac{\partial V}{ \partial P}\right|_T \propto |t|^{-\xi}
\ee
\noi
Critical exponent $\delta$: $\delta$ governs the behavior of pressure relative to the volume as,
\be
|p-p_c| \propto |v - v_c|^{\delta}
\ee
\\
\noi
Since the entropy is given by the area law,
\be
S = \pi r_h^2 = \left( \frac{ 3 V}{ 4 \pi} \right)^{2/3} \pi
\ee
and it is independent of the temperature $T$, it is easier to see that $C_V =0$. Hence the critical exponent $\alpha =0$.

To find $\beta$, lets first rewrite the equation of state in terms of $p$, $\rho$ and $\tau$ as,
\be \label{law2}
p = \frac{2}{ \lambda ( 1 + \lambda) \sigma^{2 + \lambda}} +  \frac{ 2 ( 2 + \lambda)}{ (1 + \lambda)} \frac{ \tau}{ \sigma}
\ee
Since $\epsilon = \sigma^3 -1$, $\sigma = ( 1 + \epsilon)^{1/3}$. By substituting $\sigma$ and $\tau = t +1$  in  eq$\refb{law2}$, one get
\be \label{pre}
p = 1 + a_1 t + a_2 t \epsilon + a_3 \epsilon^3 + \mathcal{O}( t^4) + \mathcal{O} ( t \epsilon^2)
\ee
where
\be 
a_1 = \frac{ 2 ( 2 + \lambda)}{ 1 + \lambda}; \hspace{0.5 cm} a_2 = - \frac{ 2 ( 2 + \lambda)}{ 3 ( 1 + \lambda)}; \hspace{0.5 cm} a_3 = \frac{ 2 + \lambda}{ 81}
\ee
Since during the phase transition we discussed in section() the pressure is constant, $p_l=p_s$, one can come to the  conclusion,
\be \label{samep}
1 + a_1 t + a_2 t \epsilon_l + a_3 \epsilon_l^3 = 1 + a_1 t + a_2 t \epsilon_s + a_3 \epsilon_s^3
\ee
Also, if we differentiate eq$\refb{law2}$  for a  fixed $t$,  we get,
\be \label{dpde}
\frac{d p}{ d \epsilon} = a_2 t + 3 a_3 \epsilon^2
\ee
During the phase transition, the Maxwell's equal area law applies, leading to,
\be \label{law3}
\int^{\epsilon_s}_{\epsilon_l} \epsilon dp = \int^{\epsilon_s}_{\epsilon_l} \epsilon \frac{dp}{d \epsilon} d \epsilon = 0
\ee
By substituting eq$\refb{dpde}$ into the eq$\refb{law3}$ and integrating, one obtain the following relation between $\epsilon_l$ and $\epsilon_s$ as,
\be \label{maxwell}
a_2 t ( \epsilon_s^2 - \epsilon_l^2) + \frac{3}{2} a_3 ( \epsilon_s^4 - \epsilon_l^4) =0
\ee
By combining eq$\refb{samep}$ and eq$\refb{maxwell}$, one can  conclude that
\be
\epsilon_l =  -\epsilon_s = \sqrt{ \frac{ -a_2 t}{ a_3}}
\ee
Now the behavior of the order parameter $\eta$ can be expanded as,
\be
\eta = V_c ( \epsilon_l - \epsilon_s) = 2 V_c \epsilon_l = 2 V_c \sqrt{ \frac{ a_2}{a_3}} \sqrt{-t} \propto |-t|^{1/2}
\ee
Hence, $\beta = \frac{1}{2}$.

To calculate the exponent $\xi$, first we will use the identity $p = \frac{P}{P_c}$ and $\epsilon = \frac{V}{V_c}- 1$ to obtain,
\be
- \left. \frac{\partial P}{ \partial V}\right|_T = \left(\frac{ \partial p}{ \partial \epsilon} \right) \frac{P_c}{V_c}
\ee
Hence,
\be
 \left. \frac{\partial V}{ \partial P}\right|_T = \frac{V_c}{P_c} \frac{1}{ \left. \frac{\partial p}{ \partial \epsilon}\right|_t} = \frac{V_c}{P_c} \frac{1}{ a_2 t}
 \ee
Now one can compute $\kappa_T$ as,
\be
\kappa_T = - \left.\frac{1}{V} \frac{\partial V}{ \partial P}\right|_T \propto \frac{1}{ a_2 P_c} t^{-1}
\ee
Hence the critical exponent $\xi =1$.

On the curve of critical isotherms, $T= T_c$ and $t =0$. Hence from eq$\refb{pre}$
\be
p \approx 1 + a_3 \epsilon^3 \ra p-1 = a_3 \epsilon^3
\ee
Therefore the critical exponent $\delta =3$.

From the above calculation of critical exponents for the AdS black hole in massive gravity, it is clear that they are the same as for the Reissner-Nordstrom-AdS black hole obtained in \cite{mann}.

 
 \section{ Specific heat capacities and local stability of the black holes}

In understanding the local stability of a black hole it is important to study the specific heat of the black hole.  The specific heat at constant pressure $P$ is given by,
\be
C_P =  T \left.  \frac{\partial S}{ \partial T}\right|_P = \frac{ 2 S \left( 8 P S^{\frac{2 + \lambda}{2}} + S ^{\frac{\lambda}{2}} + \pi^{\frac{\lambda}{2}} Q^2 \gamma ( -1 + \lambda)\right) } { \left( 8 P S^{\frac{2 + \lambda}{2}} - S ^{\frac{\lambda}{2}} - \pi^{\frac{\lambda}{2}} Q^2 \gamma ( -1 + \lambda^2)\right) }
\ee
The black hole is locally stable if $C_P >0$. In the next section we will discuss $C_P$ and local stability of black holes for the two values of $\gamma$.


 \subsection{ Specific heat  for $\gamma =-1$}

\begin{figure} [H]
\begin{center}
\includegraphics{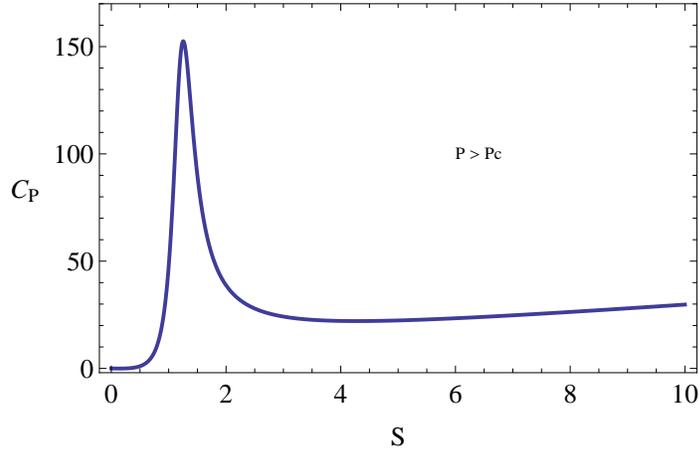}
\caption{The figure shows  $C_p$ vs $S$ for $ \gamma = -1$ for  $ P >P_c$. Here  $\lambda = 3.02, Q = 0.11$ and $P = 0.063$. For the parameters chosen $P_c = 0.0605$}
\label{cpabove}
 \end{center}
 \end{figure}

We can relate the behavior of $C_P$ to the values of $P$ relative to $P_C$. When $P > P_c$, there are no singular points for $C_P$ and it is positive except for small values of $S_c$ or small black holes. This behavior is demonstrated in Fig$\refb{cpabove}$.

\begin{figure} [H]
\begin{center}
\includegraphics{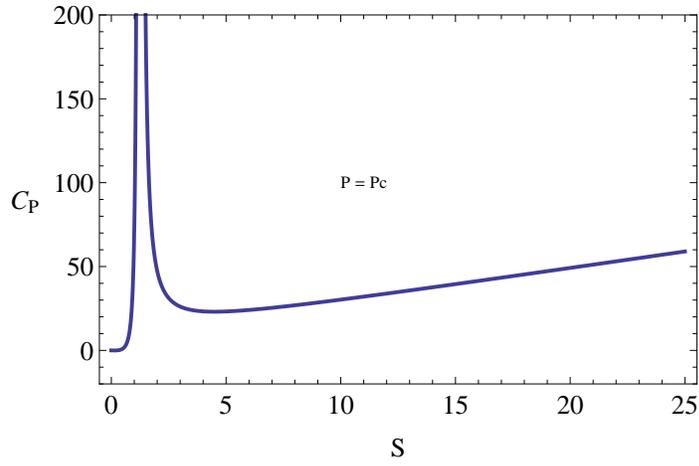}
\caption{The figure shows  $C_p$ vs $S$ for $ \gamma = -1$ for  $ P =P_c$. Here  $\lambda = 3.02, Q = 0.11$ and $P_c = 0.0605$}
\label{cpcritical}
 \end{center}
 \end{figure}

\begin{figure} [H]
\begin{center}
\includegraphics{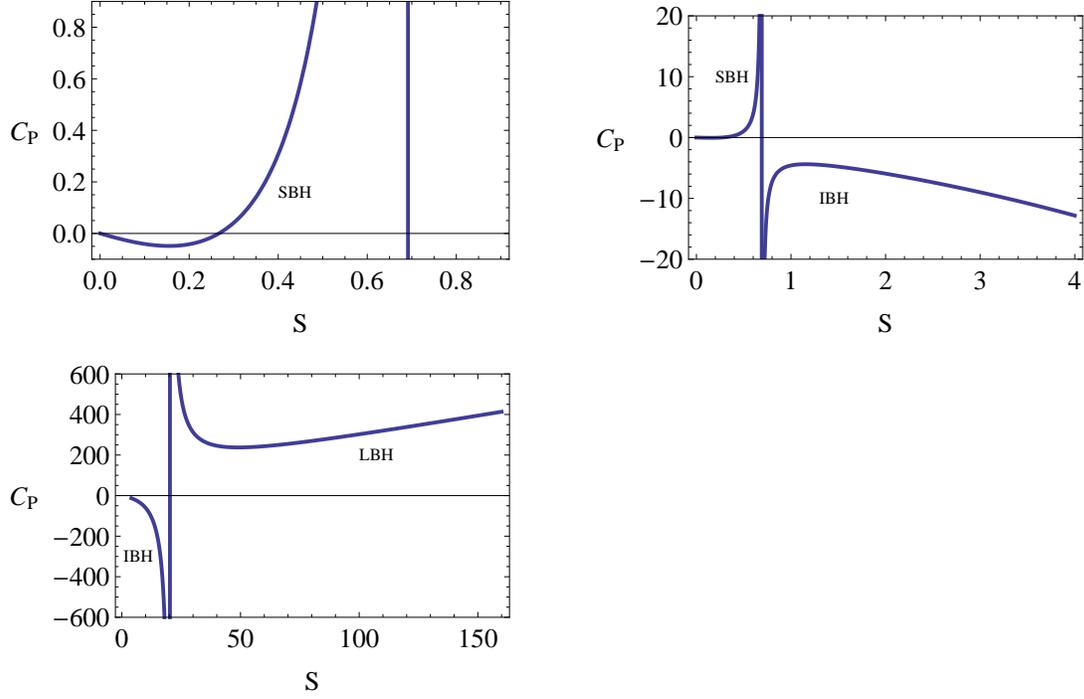}
\caption{The figure shows  $C_p$ vs $S$ for $ \gamma = -1$ for  $ P <P_c$. Here  $\lambda = 3.02, Q = 0.11$ and $P = 0.00613$. For the parameters chosen $P_c = 0.0605$}
\label{cpbelow}
 \end{center}
 \end{figure}

When $P= P_c$, $C_p$ has  a singular point at the critical value $S_c$ as shown in Fig$\refb{cpcritical}$. When the pressure is lower than than $P_c$, there are two singular pints for $C_p$ as shown in Fig$\refb{cpbelow}$. The two singular points corresponds to the maxima and the minima of the $P$ vs $r_h$ graph in Fig$\refb{pvsr1}$ for $T < T_c$. From the graph, one can observe that for small black holes (SBH) $C_P >0$ and hence they are stable. For intermediate black holes (IBH), $C_P <0$ and hence they are unstable. $C_P >0$ for large black holes (LBH) and  they are stable.

In closer observation, one can see that the denominator of $C_P$ is zero at the critical point. To observe this one can substitute the values of $P_c$, $S_c$ given in eq.$\refb{pc}$ and eq$\refb{sc}$ to $C_P$ and it will diverge.

\subsection{ Specific heat for $\gamma =1$}

\begin{figure} [H]
\begin{center}
\includegraphics{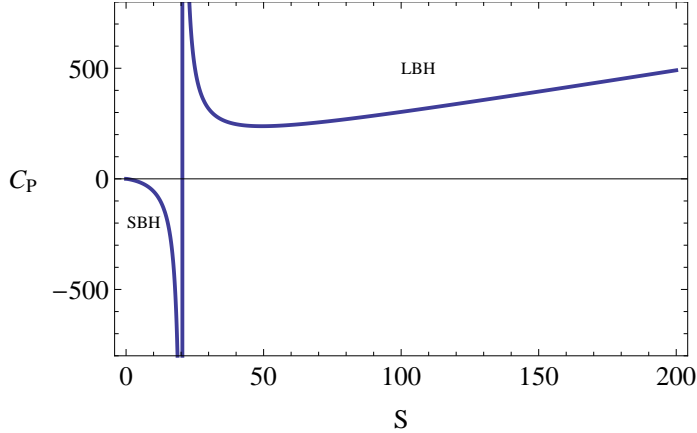}
\caption{The figure shows  $C_p$ vs $S$ for $ \gamma = 1$.  Here  $\lambda = 3.02, Q = 0.11$ and $P = 0.00613$}
\label{cpgamma}
 \end{center}
 \end{figure}

For $\gamma =1$, the $C_P$ vs $S$ is plotted in Fig$\refb{cpgamma}$. Here, $C_P <0$ until it reach a singular point. This singular point is the place where the pressure reach a maximum in the graph $P$ vs $r$ for $\gamma =1$, 
Fig$\refb{pvsr2}$.  $C_P >0$ after the singular point. Hence the small black holes are unstable and large black holes are stable.

\section{Gibbs free energy and the global stability}

Global stability of black holes can be understood by studying the corresponding  free energy. In the extended phase space, the mass of the black hole is considered as the enthalpy, not the internal energy; for a fixed charge $Q$, the corresponding free energy, Gibbs free energy $G$ is given by,
\be
 G(Q, P) = H = T S  = M - T S = \frac{r}{4} - \frac{ \gamma Q^2} {4  r^{\lambda -1}} ( 1 + \lambda) - \frac{ 2 \pi P r^3}{ 3}
\ee
We can also write $G$  in terms of $r_h, T$ and $Q$ as,
\be
G( Q, T) = \frac{1}{6} \left( 2 r - 2 \pi r^2 T - Q^2 r^{1 - \lambda} \gamma ( 2 + \lambda) \right)
\ee
The reason to write it in terms of $T$ instead of $P$  is to demonstrate how $G$ behaves when we change $P$  keeping $T$ constant.

\subsection{Phase transitions for $\gamma =-1$}

\begin{figure} [H]
\begin{center}
\includegraphics{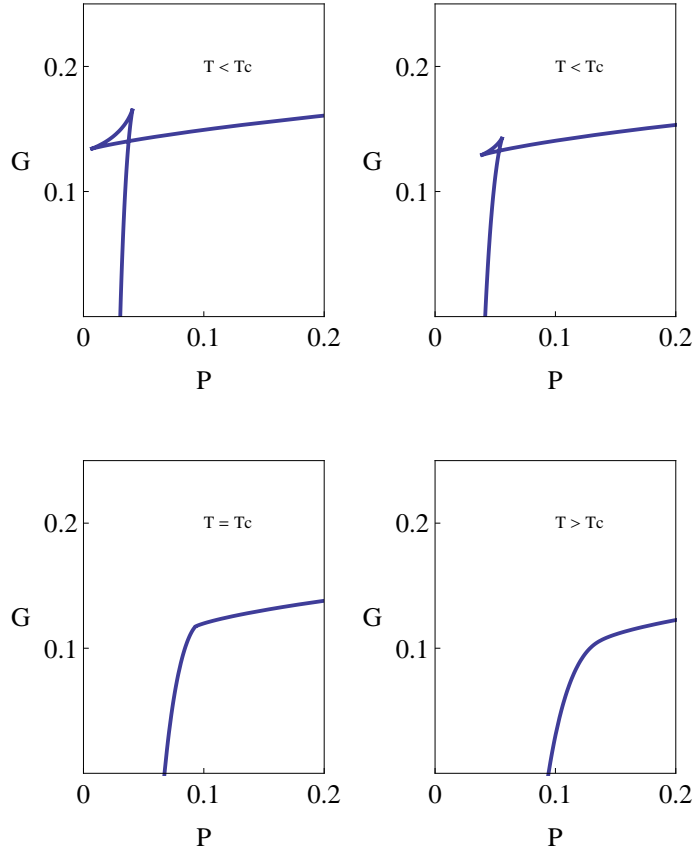}
\caption{The figure shows  $G$ vs $P$ for $ \gamma = -1$ for varying temperature. Here  $\lambda = 4.25, Q = 0.0368$ and $T_c = 0.2391$.}
\label{free3}
 \end{center}
 \end{figure}

\begin{figure} [H]
\begin{center}
\includegraphics{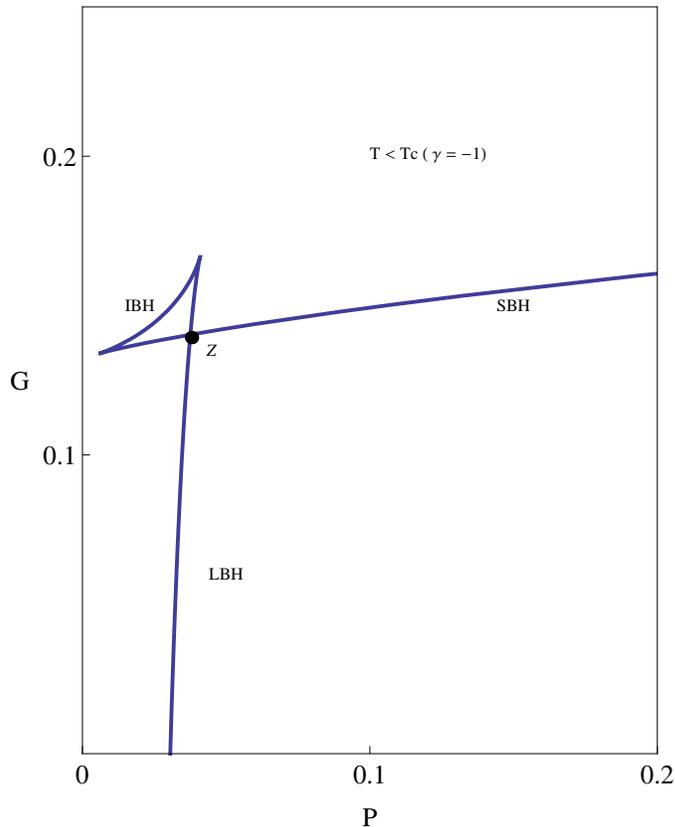}
\caption{The figure shows  $G$ vs $P$ for $ \gamma = -1$ when $T < T_c = 0.2391$. Here $T = 0.161, \lambda = 4.25$, and $ Q = 0.0368$.}
\label{free4}
 \end{center}
 \end{figure}

The Gibbs free energy is plotted against  pressure $P$ in the Fig$\refb{free3}$ for $\gamma = -1$. Here, one can see that when $T < T_c$ there is a swallow tail behavior which affirms the first order phase transition we saw in the $P$ vs $r_h$ diagram. In the first graph of Fig$\refb{free3}$, there are three branches of the swallow tail. These three branches corresponds to the small, intermediate and large black holes. When the temperature is increased, the swallow tail behavior disappears. The swallow tail behavior of the graphs are better represented in Fig$\refb{free4}$.  For high pressures, small black hole is preferred thermodynamically. When the pressure is decreased, at point $Z$, and at smaller pressures the large black hole becomes the preferred thermodynamical state. This is due to the fact that the large black holes has smaller free energy compared to the smaller black holes. Hence at point $Z$, there is small-black hole/large-black hole phase transition.

The surfaces of small and large black holes differ; there is discontinuity in the black hole areas and hence of entropy at point $Z$. This means there is release of latent heat at the point $Z$. The phase transition at point $Z$ is first order.

\subsection{ Phase transitons  for $\gamma =1$ }

\begin{figure} [H]
\begin{center}
\includegraphics{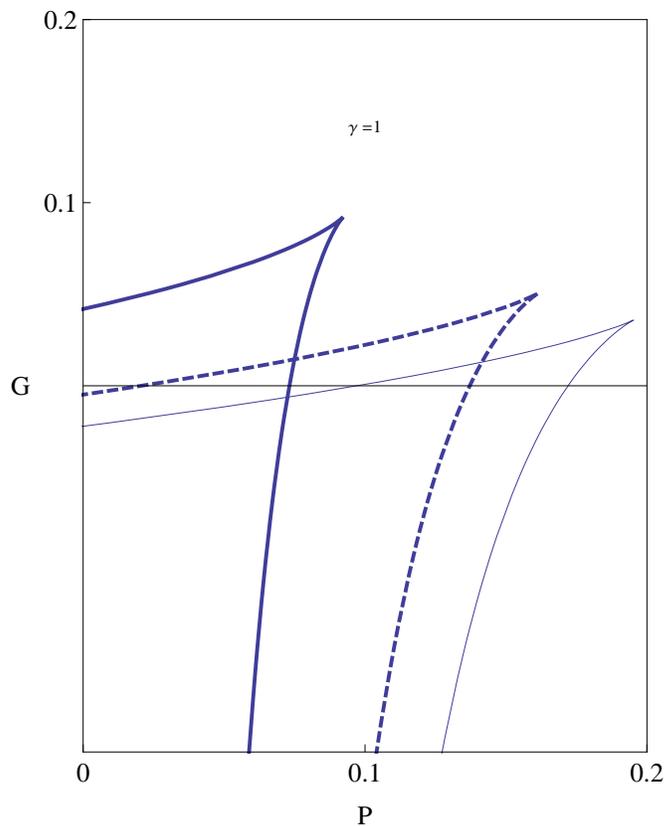}
\caption{The figure shows  $G$ vs $P$ for $ \gamma = 1$ for varying temperature. Here  $\lambda = 3.87$ and $ Q = 0.0711$ The larger the temperature the height of the singular point of the graph is lower.}
\label{free2}
 \end{center}
 \end{figure}

In the Fig$\refb{free2}$, $G$ vs  $P$ is plotted. On e can see that there are two branches meeting at a cusp like point. Wheh the tempearture is increased the free energy becomes completely negative.

\begin{figure} [H]
\begin{center}
\includegraphics{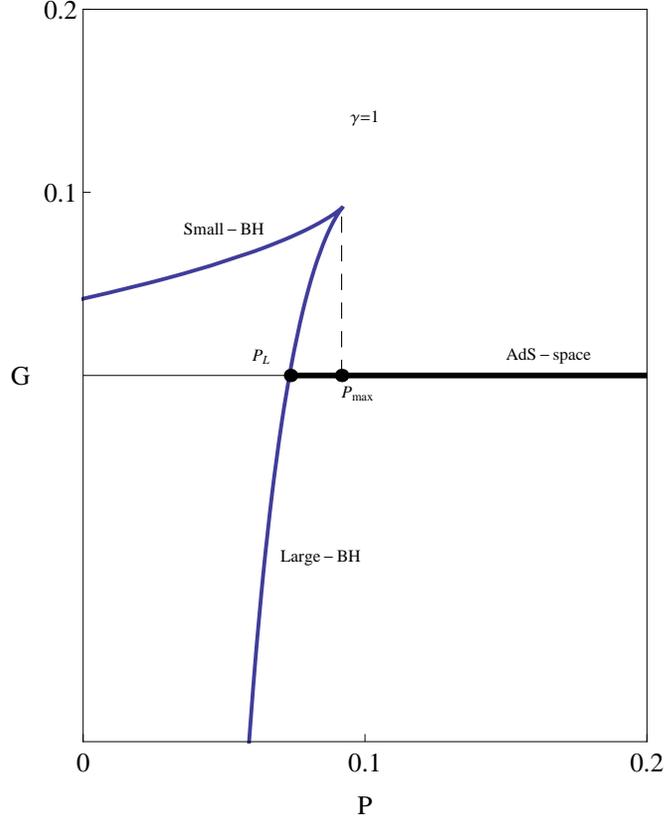}
\caption{The figure shows  $G$ vs $P$ for $ \gamma = 1$. Here $T = 0.249, \lambda = 3.87$, and $ Q = 0.0711$.}
\label{free1}
 \end{center}
 \end{figure}

We have plotted one of the graphs in detail in Fig$\refb{free1}$. When $ P > P_{max}$, no black hole can exist. Hence for that range of pressures, the thermal AdS state is preferred thermodynamically. When $P  < P_{max}$, there are two branches of black holes to choose from. The upper branch corresponds to small black holes with negative specific heat and are thermodynamically unstable. The lower branch corresponds to large black holes and are thermodynamically stable due to the positive specific heat. However, beyond the point $P_{max}$, the free energy of both branches are positive, hence, the thermal AdS state is globally preferred thermodynamic state until the point $P_L$. For pressures smaller than $P_L$, the free energy of the large black hole is negative and would be the preferred thermodynamical state. Therefore, there is a first order phase transition between thermal AdS space and the large black holes at point $P_L$. This phase transition is similar to the first order phase transition observed in the Schwarzschild-anti-de Sitter black holes \cite{page}. The value of $P_L$ can be found by solving $G =0$ for a given temperature as,
\be
P_L = \frac{ 6 \pi r_L T ( 1 + \lambda) - 3 \lambda}{ 8 \pi r_L^2 ( 2 + \lambda)}
\ee
The value of $P_{max}$ is given in eq$\refb{pmax}$.

\section{ Conclusion}

In this paper, we have studied the massive black hole in AdS space in the extended phase space. Here we have treated the pressure, P, as $P = -\frac{\Lambda}{ 8 \pi}$. The black hole system is considered to be in the canonical ensemble where the scalar charge $Q$ is taken as constant. Thermodynamical quantities such as temperature, pressure, specific heat capacity and gibbs energy is studied in detail to understand the local and global thermodynamical  stability.

There are two different types of black holes for the values of $\gamma$ in the theory. As a result, the thermodynamical quantities differ significantly for $\gamma =1$ and $\gamma =-1$.

For $\gamma =1$, the AdS black hole in massive gravity behaves similar to the Schwarzschild-anti-de Sitter black hole. The temperature has a minimum and the pressure has a maximum. There are no critical values for pressure etc. The specific heat capacity $C_P <0$ for small black holes and $C_P >0$ for large black holes. Hence small black holes are unstable locally and large black holes are stable locally. Globally, for large pressure values, the thermal AdS state is preferred  over the black holes. When the pressure is lowered, for a specific value of pressure, where the Gibbs free energy becomes zero, the large black holes take over as the preferred thermodynamic state. Hence there is a first order phase transition at that point.

For $\gamma =-1$, the thermodynamical behavior of the black holes are much   more complex. For higher temperatures, the black holes behave more like an ideal gas. There is a critical temperature where there are inflection points and critical pressure. When the pressure is lowered beyond $T_c$, Van der Waals like phase transitions occur between the small and the large black holes. The Gibbs free energy verify these phase transitions with a swallow tail type behavior. This phase transition is first order.  The values of $C_P >0$ of small and large black holes and $C_P <0$ for intermediate black holes. Hence, locally, the small and large black holes are thermodynamically stable. However, at high pressures, the small black holes are preferred and for small pressures, the large black holes are preferred. Critical exponents are computed to understand the behavior of physical quantities around the critical values. The critical exponents are the same for the Reissner-Nordstrom-AdS black holes as well as the Van der Waals liquid/gas system.

The current author also studied the thermodynamics of the same black hole in the non-extended phase space \cite{fernando1}. There the pressure was not a part of the thermodynamical quantities. There the temperature showed critical behavior when the charge $Q$ was varied. The two approaches confirm the existence of phase transitions of first order of the Van der Waals kind.

In this paper we have considered the system in the canonical ensemble where  the scalar charge $Q$ constant in analyzing the thermodynamics. One could ask the question if there would be critical behavior exists if the system is considered in the grand canonical ensemble with the potential $\Phi$ kept constant instead.  For constant $\Phi$, the equation of state becomes,
\be
P = -\frac{1}{ 8 \pi r_h^2} + \frac{T}{ 2 r_h}  + \frac{ \gamma ( 1 - \lambda)}{ 8 \pi} \frac{ \Phi^2}{ r^{ 4 - \lambda}}
\ee
We have noticed that criticality indeed can occur  for certain parameters. This is in contrast to Reissner-Nordstrom-AdS black hole where it was shown that criticality cannot happen \cite{mann}. It would be interesting to do a detail study in the grand canonical ensemble with the above equation of state.

\vspace{0.5 cm}




\begin{thebibliography}{99}


\bib{page} S. W. Hawking \& D.N. Page, {\it Thermodynamics of black holes in anti-de Sitter space}, Comm. Math. Phys. {\bf 87} 577 (1983)

\bib{witten1} E. Witten, {\it Anti-De Sitter space and holography},  Adv. Theor. Math. Phys. {\bf 2} 253 (1998)

\bib{witten2} E. Witten, {\it Anti-de Sitter space, thermal phase transition, and confinement in gauge theories}, Adv. Theor. Math. Phys. {\bf 2} 505 (1998)

\bib{cham1}  A. Chamblin, R. Emparan, C. V. Johnson \& R. C. Myers, {\it Charged AdS black holes and catastrophic holography}, Phys. Rev. {\bf D 60} 064018 (1999) 

\bib{cham2} A. Chamblin, R. Emparan, C. V. Johnson \& R. C. Myers, {\it Holography, thermodynamics, and fluctuations of charged AdS black holes}, Phys. Rev. {\bf D 60} 104026 (1999)

\bib{fernando1} S. Fernando, {\it Thermodynamics of Born-Infeld-anti-de Sitter black  holes in the grand canonical ensemble}, Phy. Rev. {\bf D 74} 104032 (2006)


\bib{juan} J. M. Maldacena, {\it The large N limit of superconformal field theories and supergravity}, Adv. Theor. Math. Phys. {\bf 2}, 231 (1988)



\bib{kastor} D. Kastor, S. Ray, \& J. Traschen, {\it Enthalpy and mechanics of AdS black holes}, Class. Quant. Grav. {\bf 26} 195011 (2009)

\bib{dolan} B. P. Dolan, {\it The cosmological constant and black hole thermodynamic potential}, Class. Quant. Grav. {\bf 28} 125020 (2011)


\bib{mann}  D. Kubiznak \& R. B. Mann, {\it P-V criticality of charged AdS black holes},  JHEP 1207:033, (2012)

\bib{mann2} R. A. Henninger \& R. B. Mann, {\it Reetrant phase transitions and van der Waals behavior for hairy black holes}, Entropy, {\bf 12} 8056 (2015)

\bib{mann3} N. Altamirano, D. Kubiznak, R. B. Mann \& Z. Sherkatghanad, {\it Thermodynamics of rotating black holes and black rings: phase transitions and thermodynamic volume}, Galaxies {\bf 2} 89 (2014)

\bib{mann4} S. Gunasekaran, R. B. Mann, \& D. Kubiznak, {\it Extended phase space thermodynamics for charged and rotating black holes and Born-Infeld vacuum polarization}, JHEP 11: 10 (2012)

\bib{mann5} A. Rajagopal, D. Kubiznak, \& R. B. Mann, {\it Van der Waals black hole}, Phys. Lett. {\bf B 737} 277 (2014)

\bib{cao}  J. Xu, L. Cao, \& Y. Hu, {\it P-V criticality in the extended phase space of black holes in massive gravity}, Phys. Rev. {\bf D 91} 124033 (2015)

\bib{liu2}  J. Mo \& W. Liu, {\it P-V criticality of topological black holes in Lovelock-Born-Infeld gravity}, Eur. Phys. Jour. {\bf C 74} 2836 (2014)

\bib{hendi} S. H. Hendi, S. Panahiyan, \& B. E. Panah, {\it Extended phase space thermodynamics and P-V criticality of black holes with Born-Infeld type nonlinear electrodynamics}, Int. Jour. Mod. Phys. {\bf D 25} 1650010 (2016)

\bib{hendi2} S. Hendi, \& M. H. Vahidinia, {\it Extended phase space thermodynamics of black holes with nonlinear source}, Phys. Rev. {\bf D88} 084045 (2012)

\bib{mo}   J. Mo, G. Li, \& X. Xu, {\it Effects of power-law Maxwell field on the Van der Waals like phase transition of higher dimensional dilaton black holes}, Phys. Rev.{\bf D 93} 084041 (2016)

\bib{zhang} M. Zhang \& W Liu, {\it Coexistent physics of massive black holes in the phase transitions}, arXiv: 1610.03648

\bib{li} G. Li, {\it Effects of dark energy on P-V criticality of charged AdS black-holes}, Phys. Lett. {\bf B735} 256 (2014)

\bib{mo2} J. Mo, G. Li, \& X. Xu, {\it Combined effects of f(R) gravity and conformally invariant Maxwell field on the extended phase space thermodynamics of higher-dimensional black holes}, Eur. Phys. Jour. {\bf C 76} 545 (2016)

\bib{cai} R Cai, L. Cao, \& R. Yang, {\it P-V criticality in the extended phase space of Gauss-Bonnet black holes in AdS space} JHEP:1309 005 (2013)

\bib{cai2} R.  Cai, Y. Hu, Q. Pan, \& Y. Zhang, {\it Thermodynamics of black holes in massive gravity}, Phys. Rev. {\bf D 91} 024032 (2015)

\bib{azg} M. Azreg-Ainou, {\it Black hole thermodynamics: No inconsistency via the inclusion of the missing P-V terms} , Phys. Rev. {\bf D 91}, 064049 (2015)

\bib{pou} J. Sadeghi, B. Pourhassan, \& M. Rostami, {\it P-V criticality of logarithmic corrected dyonic charged AdS black hole}, Phys. Rev. {\bf D 94} 064006 (2016)



\bib{ligo}  B. P. Abbott et.al., { \it Observation of gravitational waves by a binary black hole merger}, Phys. Rev. Lett. {\bf 116} 061102 (2016)

\bib{pauli} M. Fierz  \& W. Pauli, {\it On relativistic wave equations for particles of arbitrary spin in an
electromagnetic field},  Proc. R. Soc. London, Ser. {\bf A 173}  211 (1939)


\bib{boul}  D. G. Boulware \& S. Desrer, {\it Can gravitation have a finite range ?}, Phys. Rev. {\bf D 6} 3368 (1972)


\bib{drgt1}  C. de Rham \&  G. Gabadadze, {\it Generalization of the Fierz-Pauli action}, Phys. Rev. {\bf D 82} 044020 (2010)

\bib{drgt2} C. de Rham \&  G. Gabadadze \& A. J. Tolley, {\it  Resummation of massive gravity}, Phys. Rev. Lett. {\bf 106} 231101 (2011)

\bib{dgp} G. Dvali, G. Gabadadze \& M. Porrati, {\it 4D gravity on a brane in 5D Minkowski space}, Phys. Lett. {\bf B 485} 208 (2000)

\bib{town} E. A. Bergshoeff, O. Hohm, \& P. K. Townsend, {\it Massive gravity in three dimensions}, Phys. Rev. Lett. {\bf 102} 201301 (2009)

\bib{claudia} C. de Rham, {\it Massive gravity}, Living Rev. Relativity {\bf 17} 7 (2014)

\bib{kurt} K. Hinterbichler, {\it Theoretical aspects of massive gravity}, Rev. Mod. Phys. {\bf 84} 671 (2012)


\bib{dubo} S. L. Dubovsky, {\it Phases of massive gravity}, JHEP {\bf 0410} 076 (2004)

\bib{ruba2} V. A. Rubakov \&  P. G. Tinyakov, {\it Infrared-modified gravities and massive gravitons}, Phys. Usp. {\bf 51} 759  (2008)

\bib{luty} N. Arkani-Hamed, H. Cheng, M.A. Luty \&  S. Mukohyama, {\it Ghost condensation and a consistent infrared modification of gravity}, JHEP {\bf 0405} 074 (2004)

\bib{ruba} V. Rubakov, {\it Lorentz-violating graviton masses: getting around ghosts, low strong coupling scale and VDVZ discontinuity}, hep-th/0407104.

\bib{pilo1} D.Blas, D. Comelli, F. Nesti, \&  L. Pilo,  {\it Lorentz Breaking Massive Gravity in Curved Space}, Phys. Rev. {\bf D80} 044025 (2009)



\bib{tinya}  M. V. Bebronne \& P. G. Tinyakov, {\it Black hole solutions in massive gravity}, JHEP 0904:100, 2009; Erratum-ibid.1106:018, (2011)

\bib{pilo} D. Comelli, F. Nesti \& L. Pilo, {\it Stars and (Furry) black holes in Lorentz breaking massive gravity}, Phys. Rev. {\bf D 83} 084042 (2011)


\bib{fernando1} S. Fernando, {\it Phase transitions in black holes in massive gravity},  Mod. Phys. Lett.{\bf A 31} 1650096 (2016)

\bib{fernando2} S. Fernando \& T. Clark, {\it Black holes in massive gravity: quasinormal modes of scalar perturbations}, Gen. Rel. Grav. {\bf 46 }   1834 (2014)

\bib{fernando3} S. Fernando, {\it Black holes in massive gravity: quasinormal modes of Dirac field  perturbations},  Mod. Phys. Lett.{\bf A 30}	1550147 (2015)

\bib{capela} F. Capela \& G. Nardini, {\it Hairy black holes in massive gravity: Thermodynamics and phase structure}, Phys. Rev. {\bf D 86} 024030 (2012)

\bib{mirza} B. Mirza \& Z. Sherkatghanad, {\it Phase transitions of hairy black holes in massive gravity and thermodynamics behavior of charged AdS black holes in an extended phase space}, 
Phys. Rev. {\bf D 90} 084006 (2014)

\bib{and}  A. Addazi, \& S. Capozziello, {\it External stability for spherically symmetric solutions in Lorentz breaking massive gravity}, Int. Jour. Theo. Phys. {\bf 54} 1818 (2015)


\bib{hill} T. L. Hill, {\it An introduction to statistical thermodynamics}, Dover Publications, Inc, New York, (1986)


\bib{book1}  N. Goldenfeld, {\it Lectures on phase transitions and the renormalization group}, Addison Wesley, (1992)

\end{thebibliography}
\end{document}